\documentclass[aps,prb,amsmath,amssymb,superscriptaddress, reprint]{revtex4-1}

\usepackage{graphicx}

\begin{document}

\title{Correlation-enhanced control of wave focusing in disordered media}

\author{Chia Wei Hsu}
\affiliation{Department of Applied Physics, Yale University, New Haven, Connecticut 06520, USA}
%\email{chiawei.hsu@yale.edu}

\author{Seng Fatt Liew}
\affiliation{Department of Applied Physics, Yale University, New Haven, Connecticut 06520, USA}

\author{Arthur Goetschy}
\affiliation{ESPCI ParisTech, PSL Research University, CNRS, Institut Langevin, 1 rue Jussieu, F-75005 Paris, France}

\author{Hui Cao}
\affiliation{Department of Applied Physics, Yale University, New Haven, Connecticut 06520, USA}

\author{A. Douglas Stone}
\affiliation{Department of Applied Physics, Yale University, New Haven, Connecticut 06520, USA}

\begin{abstract}
A fundamental challenge in physics is controlling the propagation of waves in disordered media despite strong scattering from inhomogeneities.
Spatial light modulators enable one to synthesize (shape) the incident wavefront, optimizing the multipath interference to achieve a specific behavior such as focusing light to a target region. 
However, the extent of achievable control was not known when the target region is much larger than the wavelength and contains many speckles.
Here we show that for targets containing more than $g$ speckles, where $g$ is the dimensionless conductance, the extent of transmission control is substantially enhanced by the long-range mesoscopic correlations among the speckles.
Using a filtered random matrix ensemble appropriate for coherent diffusion in open geometries, we predict the full distributions of transmission eigenvalues as well as universal scaling laws for statistical properties, in excellent agreement with our experiment.
This work provides a general framework for describing wavefront-shaping experiments in disordered systems.
\end{abstract}

\maketitle

Waves propagating through a disordered medium undergo multiple scattering from the inhomogeneities.
Interference among the multiply scattered fields has important consequences that cannot be described with incoherent diffusion~\cite{Akkermans_book, Sheng_book}.
By controlling the incident wave (``wavefront shaping,'' WFS), one can manipulate this interference and drastically modify the transport of light, microwaves, and acoustic waves~\cite{2012_Mosk_nphoton}.
% 2015_Vellekoop_OE_review
One early and notable example is focusing light onto a {\it local} speckle-sized target through aligning the scattered fields there~\cite{2007_Vellekoop_OL, 2008_Yaqoob_nphoton, 2010_Hsieh_OE, 2012_Conkey_OE},
which has led to advances in imaging within biological tissue and other scattering materials~\cite{2015_Horstmeyer_nphoton}.
% 2010_Vellekoop_OL, 2012_Wang_ncomms, 2012_Katz_nphoton, 2012_Si_nphoton, 2013_Judkewitz_nphoton, 2015_Kang_nphoton
The transport through disordered structures is described by a random field transmission matrix, and the use of WFS over such local properties has treated the matrix elements as having only short-range correlations on the scale of a single speckle~\cite{2007_Vellekoop_OL, 2008_Sprik_PRB, 2009_Aubry_PRL, 2010_Popoff_PRL, 2011_Popoff_NJP, 2015_Dremeau_OE}.
However, it has long been known 
that diffusive waves also exhibit long-range and infinite-range correlations~\cite{1987_Stephen_PRL, 1988_Feng_PRL, 1988_Mello_PRL, 2002_Sebbah_PRL};
this was previously noted in the context of electron transport through mesocopic structures, where correlations lead to anomalously large conductance fluctuations~\cite{1987_Lee_PRB}.
% Akkermans_book_Ch12
% 1994_Berkovits_PR, 2001_Pnini_chapter, 2001_Scheffold_chapter, Sheng_book
The long-range correlations are related to the existence of near-unity-transmission input states (``open channels'')~\cite{1984_Dorokhov_SSC, 1986_Imry_EPL, 1988_Mello_AP, 1994_Nazarov_PRL, 2014_Gerardin_PRL, 2016_Sarma_PRL},
% 2014_Popoff_PRL
and have measurable effects on other {\it global} statistical properties of diffusive waves such as the total transmission variance~\cite{1992_deBoer_PRB, 1997_Stoytchev_PRL, 1997_Scheffold_PRB, 2013_Strudley_nphoton},
% 1994_deBoer_PRL
% 1990_Genack_PRL, 1995_Nieuwenhuizen_PRL, 2014_Strudley_OL
the increased background for maximally focused waves~\cite{2008_Vellekoop_PRL, 2012_Davy_PRB, 2013_Davy_OE, 2016_Ojambati_PRA},
and the singular values of large transmission matrices~\cite{2012_Shi_PRL, 2013_Yu_PRL, 2016_Yu_PRB, 2016_Akbulut_PRA}.
% 2013_Akbulut_thesis
% theory: 1989_Pnini_PRB, 2011_Choi_PRB, 2013_Goetschy_PRL, 2014_Liew_PRB, 2015_Davy_PRL, 2015_Hsu_PRL, 2016_Yamilov_PRB
With the rapid growth of WFS, an important question, both scientifically and technologically, is how correlations affect the coherent control over targets larger than a single speckle and smaller than the full transmitted pattern, {\it i.e.}~in between local and global.
This intermediate regime remains poorly understood but is relevant for many applications ranging from telecommunications and cryptography to photothermal therapy and the optical or ultrasound imaging of large objects behind a scattering medium.

Here, we demonstrate the effects of correlations by means of optical WFS experiments in this interesting regime.
WFS enables dynamic control over how much light is transmitted into a given target,
and we find that for large targets, correlations increase the range of control significantly beyond what would be achievable if correlations were negligible (as is typically assumed).
Physically this is because in a multiply scattering medium, the transmitted flux is carried by roughly only $g$ open channels~\cite{1984_Dorokhov_SSC, 1986_Imry_EPL, 1988_Mello_AP, 1994_Nazarov_PRL};
here $g$ is the analog of the dimensionless conductance or Thouless number for electron transport in a waveguide, and its definition in an open geometry will be discussed below.
When the target region is large enough that the number $M_2$ of speckles in it exceeds $g$, the output channels are necessarily correlated. 
Such positive correlations reduce the effective degrees of freedom $M_2^{({\rm eff})}$ that need to be controlled and lead to the increased control range and other correlation effects.

WFS experiments on strongly scattering media are typically performed in an open geometry: the illumination spot spreads laterally as wave diffuses into the medium.
A rigorous random matrix theory for such a set-up was not known until recently, when the ``filtered random matrix" ensemble (FRM) was introduced~\cite{2013_Goetschy_PRL} and conjectured to apply to diffusion with an open boundary~\cite{2014_Popoff_PRL}.
There were some partial tests of the FRM eigenvalue distributions~\cite{2013_Yu_PRL, 2016_Akbulut_PRA, 2016_Yu_PRB}, but not with lateral diffusion accounted for rigorously.
% 2013_Akbulut_thesis
Here we show precisely how the FRM can be applied to an arbitrary open diffusion experiment, and confirm its predictions for the full distributions of transmission eigenvalues.
We prove that the ratio $M_2/g$ determines the presence or absence of significant correlation effects and derive new scaling laws for measurable statistical quantities that are found to agree very well with our experimental data.
This theoretical framework is applicable to the WFS for light as well as microwaves and acoustic waves.

When we modulate $M_1$ incident channels with a spatial light modulator (SLM) and collect $M_2$ channels transmitted into a target region, the input-output relation can be written as $|\psi_{\rm out}\rangle = \tilde{t} |\psi_{\rm in}\rangle$ where $|\psi_{\rm in}\rangle$ and $|\psi_{\rm out}\rangle$ are vectors of length $M_1$ and $M_2$ that contain the input and output field amplitudes.
Here $\tilde{t}$ is an $M_2$-by-$M_1$ transmission matrix, and we use the tilde to indicate
the exclusion of unmodulated inputs and uncollected outputs~\cite{2013_Goetschy_PRL}.
The total flux into the focal target is $T = \langle \psi_{\rm out} | \psi_{\rm out}\rangle = \langle \psi_{\rm in} | \tilde{t}^\dagger \tilde{t} |\psi_{\rm in}\rangle$; the variational principle guarantees that the maximal and minimal transmitted fluxes are the extremal eigenvalues of $\tilde{t}^\dagger \tilde{t}$. The corresponding eigenvectors are the optimal input wavefronts that the SLM should synthesize. 
Thus the variance ${\rm Var}(\tilde{\tau})$, where $\tilde{\tau}$ denotes the eigenvalues of $\tilde{t}^\dagger \tilde{t}$, is a measure of the range of focused transmission that is achievable by WFS.
If the matrix elements of  $\tilde{t}$ were uncorrelated random numbers, for sufficiently large $M_1$ and $M_2$ the eigenvalues would follow the Mar\v{c}enko--Pastur (MP) distribution~\cite{1967_MP}, which has variance ${\rm Var}(\tilde{\tau}^{({\rm MP})}) = \langle \tilde{\tau} \rangle^2 M_1/M_2$.
The ratio between ${\rm Var}(\tilde{\tau})$ and ${\rm Var}(\tilde{\tau}^{({\rm MP})})$ is a measure of how correlations affect the range of coherent control.

\begin{figure}[t]
   \includegraphics{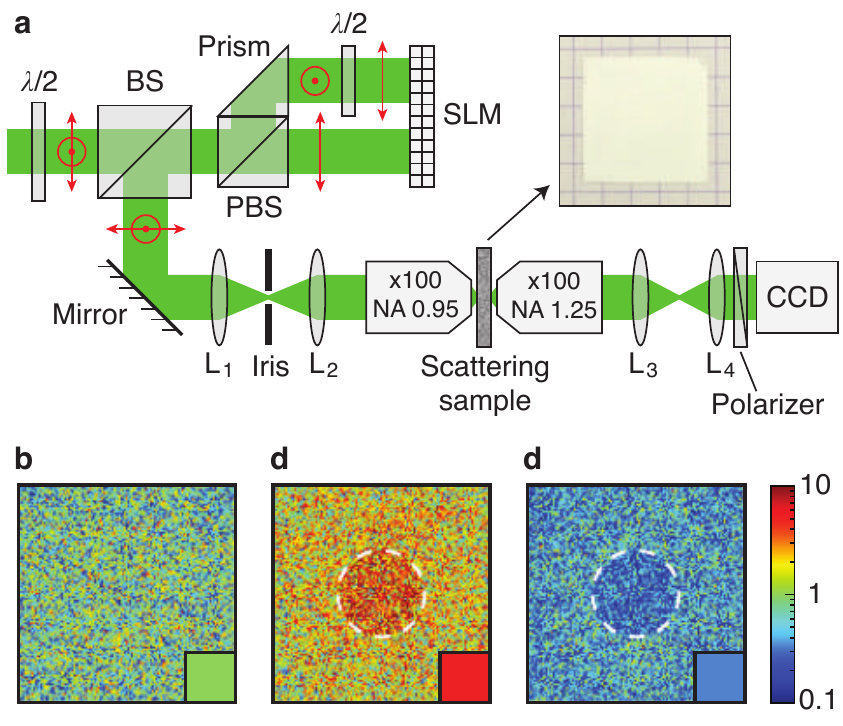} 
   \caption{
   \textbf{Experimental set-up and representative output patterns.}
   {\bf a}, A phase-only spatial light modulator (SLM) generates the desired incident wavefront and, together with a CCD camera, measures the optical transmission matrix of the scattering sample (average transmission $\bar{T} \approx$ 3\%, picture shown in the inset).
   $\lambda/2$, half-wave plate; BS, beam splitter; PBS, polarizing beam splitter; L$_{1-4}$, lenses.
   {\bf b--d}, Output pattern on the CCD for a random input wavefront ({\bf b}) and for wavefronts optimized for maximal ({\bf c}) and minimal ({\bf d}) transmission into a target region containing $M_2 \approx 1700$ speckles (within the white dashed circle), in saturated log scale; square on the corner indicates the average intensity inside the target.
   }
   \label{fig1}
\end{figure}

We study the transport through a slab of zinc oxide (ZnO) microparticles (median diameter $\approx$ 200 nm) deposited on a cover slip, with slab thickness $L \approx 60$ $\mu$m and total transmission $\bar{T} \approx$ 3\%.
The incident wavefront (wavelength $\lambda$ = 532 nm) is modulated with a phase-only SLM and then focused onto the sample with a high-NA objective, and the transmitted light is collected on a CCD camera; see the schematic illustration in Fig.~1a and details in Methods.
In our set-up, the SLM/CCD pixels modulate/detect different angles incident onto/transmitted from the sample.
The nearby SLM pixels are grouped into macropixels; smaller macropixels correspond to more finely spaced incident angles, with a larger illumination spot and more available input channels.
The illumination spot size determines the crucial parameter $g$ in an open geometry,
and we consider three macropixel sizes that correspond to illumination diameters $D_{\rm in} \approx 6, 12, 24$ $\mu$m; the number of modes we modulate is $M_1 = 128, 512, 2048$ respectively.
Our output speckle grains are slightly larger than one CCD pixel (the intensity autocorrelation width is 1.5 pixels),
and we keep only one pixel out of $2\times 2$ pixels in the data recorded on the CCD to remove correlations among neighboring pixels;
thus each remaining CCD pixel corresponds to one output channel, free of short-range correlations.
With this set-up, we measure the transmission matrix $\tilde{t}$ using a phase-shifting common-path interferometric method (Methods).
Once the transmission matrix is measured, we use it to predict the optimal phase-only wavefront for a given target (Methods), and then measure the output on the CCD when such a wavefront is applied on the SLM.
Exemplary outputs for enhanced and suppressed transmission into a large target are shown in Fig.~1b--d.

The intensity correlation function between the output channels are calculated from the measured transmission matrices and shown in Supplementary Fig.~1 [section I of the Supplementary Information (SI)].
Long-range correlations are readily seen in the data, which explain why in Fig.~1c--d, the background speckle intensities outside the target increase (or decrease) with those inside the target, as observed previously~\cite{2008_Vellekoop_PRL, 2012_Davy_PRB, 2013_Davy_OE, 2016_Ojambati_PRA}.

\begin{figure}[t]
\centerline{
   \includegraphics[scale=0.9]{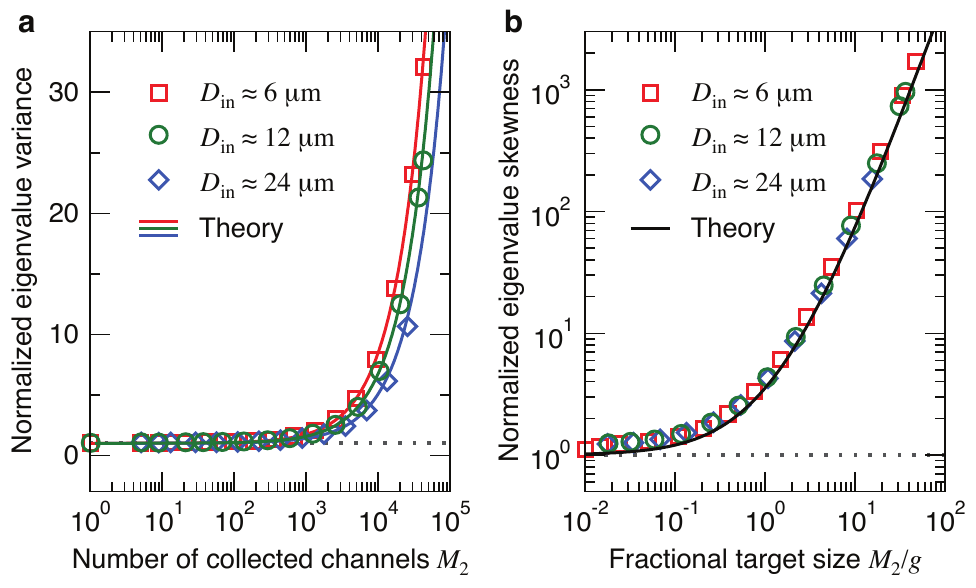} 
}
   \caption{
   \textbf{Correlation effects in wavefront shaping and their universal scaling.}
   {\bf a}, Eigenvalue variance of the measured transmission matrix divided by that of an uncorrelated matrix.
   This quantity characterizes how the correlations between matrix elements enhance the range of control in wavefront shaping, and it departs significantly from unity (dotted line) when the target size $M_2$ exceeds the dimensionless conductance $g$ (values given in the text).
   Symbols: data with different illumination diameters $D_{\rm in}$. Lines: scaling curve of equation~\eqref{eq:var_g} using the measured $g$'s.
   {\bf b}, Eigenvalue skewness of the measured transmission matrix divided by that of an uncorrelated matrix. Black line is the scaling curve of equation \eqref{eq:skew_g}.
\label{fig2}
}
\end{figure}

We obtain the eigenvalues of $\tilde{t}^\dagger \tilde{t}$ from the measured transmission matrices, for circular targets of increasing sizes.
As the target size grows, the eigenvalue variance becomes larger than the uncorrelated MP variance, as shown in Fig.~2a.
For small targets ($M_2 \lesssim 10^3$), we observe ${\rm Var}(\tilde{\tau}) \approx {\rm Var}(\tilde{\tau}^{({\rm MP})})$ with no obvious correlation effects, consistent with prior work
%on local (speckle-sized) focusing~\cite{2007_Vellekoop_OL, 2008_Yaqoob_nphoton, 2010_Hsieh_OE, 2012_Conkey_OE}
involving small transmission matrices~\cite{2008_Sprik_PRB, 2009_Aubry_PRL, 2010_Popoff_PRL, 2011_Popoff_NJP, 2015_Dremeau_OE}.
% 2012_Kim_nphoton
However, for large targets ($M_2 \gtrsim 10^3$), ${\rm Var}(\tilde{\tau})$ becomes significantly larger than ${\rm Var}(\tilde{\tau}^{({\rm MP})})$, indicating an enhanced range of control due to correlations.
The ratio between the eigenvalue range $\tilde{\tau}_{\rm max} - \tilde{\tau}_{\rm min}$ and that from an uncorrelated matrix $\tilde{\tau}_{\rm max}^{({\rm MP})} - \tilde{\tau}_{\rm min}^{({\rm MP})}$ is shown in Supplementary Fig.~2 (SI section II), which follows the same trend as the eigenvalue variance.

\begin{figure}[t]
   \includegraphics[scale=0.95]{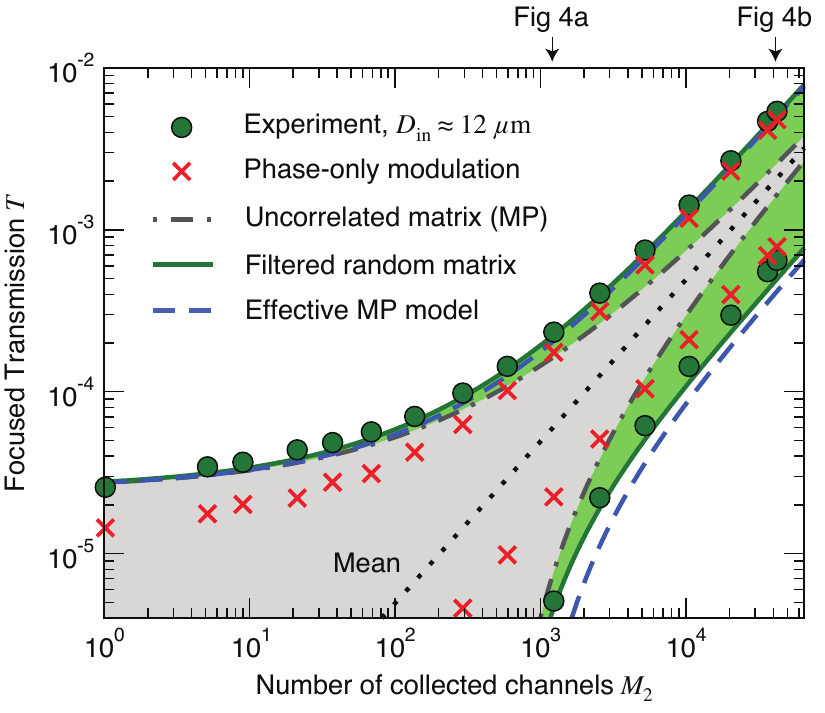} 
   \caption{
   \textbf{Allowed and achieved range of focused transmission as a function of the target size.}
   The extremal eigenvalues of the measured transmission matrix (green circles) give the allowed range.
   The achieved range through phase-only wavefront modulation (red crosses) is slightly narrower but still exceeds the Mar\v{c}enko-Pastur (MP) law for uncorrelated matrices (gray lines with shading) when the target is large.
   The allowed range is well described by theoretical predictions using the filtered random matrices (green lines with shading) and the effective MP model (blue dashed lines).
   }
   \label{fig3}
\end{figure}

The extremal eigenvalues $\tilde{\tau}_{\rm max}$ and $\tilde{\tau}_{\rm min}$ are shown in Fig.~3 (green filled symbols) for the case $D_{\rm in} \approx 12$ $\mu$m.
They stretch a wider range than the uncorrelated ones $\tilde{\tau}_{\rm max/min}^{({\rm MP})} = \langle \tilde{\tau} \rangle (1\pm\sqrt{M_1/M_2})^2$ (gray dot-dashed lines), and the difference grows with $M_2$.
We can readily achieve this extended range experimentally: when $M_2 \gtrsim 3000$, the largest and the smallest focused transmission reached with our phase-only SLM (red crosses) cover a wider range than the uncorrelated extrema, even though the uncorrelated extrema were calculated assuming both phase and amplitude modulations.
Having access to the transmission matrix is important as it helps us find near-optimal wavefronts (Methods); a recent experiment~\cite{2016_Ojambati_PRA} used feedback-based optimization and reported enhancements lower than the uncorrelated value $\tilde{\tau}_{\rm max}^{({\rm MP})} / \langle \tilde{\tau} \rangle$ because it did not reach a near-optimal wavefront.

\begin{figure}[t]
   \includegraphics{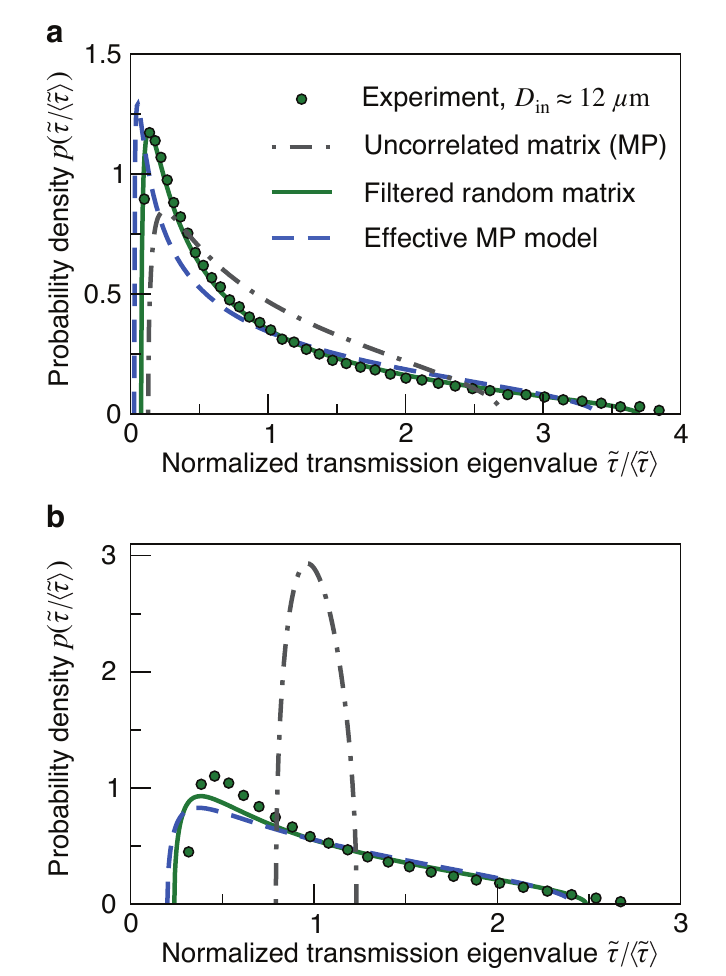} 
   \caption{
   \textbf{Full distributions of transmission eigenvalues.}
   Data are shown for the two target sizes marked in Fig.~3: {\bf a}, $M_2 \approx 10^3 \approx g$ and {\bf b}, $M_2 \approx 4 \times 10^4 \gg g$.
   The experimental data (green circles; averaged over 10 data sets) differ from the uncorrelated MP law (gray dot-dashed lines) but are described accurately by the filtered random matrix theory (green solid lines) and approximately by the effective MP model (blue dashed lines).
   }
   \label{fig4}
\end{figure}

The full distributions of eigenvalues are shown in Fig.~4 for two representative target sizes.
When $M_2 \approx 10^3$, the experimental distribution already differs detectably from the uncorrelated MP law (Fig.~4a).
With a much larger target of $M_2 \approx 4 \times 10^4$, the experimental distribution spreads five times the width of the corresponding MP distribution (Fig.~4b)---a drastic change due to correlations.

To understand and to describe quantitatively these correlation effects, we make the following ansatz:
\begin{quote}
The $M_2$-by-$M_1$ partial transmission matrix $\tilde{t}$ of an open disordered slab measured with a finite illumination area can be treated as a filtered matrix drawn from a larger $N_2$-by-$N_1$ full transmission matrix $t$ of a disordered coherent conductor in a closed waveguide of non-uniform width.
\end{quote}
The parameters of the unknown full matrix $t$ account for the effects of finite illumination area and lateral diffusion in an open geometry, and they remain to be determined.
It is known that for closed diffusive waveguides in the absence of absorption, $t^\dagger t$ has a bimodal eigenvalue density~\cite{1984_Dorokhov_SSC, 1986_Imry_EPL, 1988_Mello_AP}
\begin{equation}
\label{eq:pT}
p_{t^\dagger t}(\tau) = \frac{\bar{T}}{2 \tau\sqrt{1-\tau}}
\end{equation}
that is universal~\cite{1994_Nazarov_PRL, 2016_Yamilov_PRB}
% 2015_Jin_arXiv
and parametrized only by the average transmission $\bar{T}$;
this allows us, using the FRM, to predict the statistical properties of $\tilde{t}$.

Equation~\eqref{eq:pT} differs drastically from the MP distribution~\cite{1967_MP}, indicating that the matrix elements of $t$ are necessarily correlated.
In particular, this asymmetric bimodal distribution indicates that the transmitted waves consist mainly of a relatively small number, 
$g \ll N_1$, of ``open'' eigenchannels with order-unity transmission ($\tau \approx 1$), while most of the other eigenchannels have $\tau \approx 0$ and barely contribute to transmission.
The available degrees of freedom at the output is not the number of output channels defined by the geometry, $N_2$, but 
rather it is approximately $g \equiv \langle {\rm Tr}(t^\dagger t) \rangle = N_1 \bar{T}$;
more precisely, it can be defined by the participation number $\langle(\sum_{n=1}^{N_1}\tau_n)^2/(\sum_{n=1}^{N_1}\tau_n^2)\rangle$ (refs.~\onlinecite{2012_Davy_PRB, 2013_Davy_OE,2016_Verrier_PRB}),
which is $3g/2$ for the bimodal distribution in equation~\eqref{eq:pT}.
If we collect more output channels than the available degrees of freedom at the output (when $M_2 > 3g/2$), we expect to see strong correlation effects.

The intuitive discussion in the preceding paragraph can be made quantitative by using the analytic FRM formalism~\cite{2013_Goetschy_PRL} to describe the matrix filtering process in our ansatz.
In general, this requires knowing the three parameters $N_1,N_2,\bar{T}$ of the unknown matrix $t$, in addition to the known size $M_1, M_2$ of the measured matrix $\tilde{t}$.
But for thick samples (specifically, when $\bar{T} \ll 2/3$, which is always the case in the diffusive regime), our derivation (SI section III) shows that the variance normalized to the uncorrelated MP variance is simply
\begin{equation}
\label{eq:var_g}
\frac{{\rm Var}(\tilde{\tau})}{{\rm Var}(\tilde{\tau}^{({\rm MP})})} =  1 + \frac{2 M_2}{3g},
\end{equation}
which depends on a single parameter, $M_2/g$ 
[recall that ${\rm Var}(\tilde{\tau}^{({\rm MP})}) = \langle \tilde{\tau} \rangle^2 M_1/M_2]$.
As expected, when the target region contains more than $3g/2$ channels, the eigenvalues exhibit substantially more variation than the uncorrelated MP behavior, leading to a wider-than-expected range for coherent control.

To compare the experimental data with equation~\eqref{eq:var_g}, the only parameter we need is the dimensionless conductance $g$.
For a fixed input, the intensity correlation between far-away output speckles equals $2/(3g)$ (refs.~\onlinecite{1987_Stephen_PRL, 1988_Feng_PRL, 1988_Mello_PRL}), allowing us to determine $g$ from the experimental correlation functions (Supplementary Fig.~1).
We also determine $g$ through the measured variance of the normalized total transmission (as in refs.~\onlinecite{1992_deBoer_PRB, 1997_Scheffold_PRB}), which also equals $2/(3g)$.
The two methods yield almost the same values of $g$, whose average values are $g = 894 \pm 26, 1164 \pm 38, 1642 \pm 82$ for the three illumination sizes $D_{\rm in} \approx 6, 12, 24$ $\mu$m considered here (SI section IV).
Analytic expressions of $g$ for an open geometry~\cite{1989_Pnini_PRB, 1992_deBoer_PRB, 1997_Scheffold_PRB, 2002_Garcia-Martin_PRL, 2014_Popoff_PRL} predict similar values, which we describe in SI section V.

With $g$ determined, we compare equation~\eqref{eq:var_g} to the eigenvalue variance without any free parameter and observe excellent quantitative agreement (Fig.~2a).
Equation~\eqref{eq:var_g} reveals that the MP-normalized eigenvalue variance follows a scaling law with respect to a single-parameter $M_2/g$.
Furthermore, we show in SI section III that when $\bar{T} \ll 4/15$, the third central moment ${\rm Skew}(\tilde{\tau}) \equiv \langle (\tilde{\tau} - \langle \tilde{\tau} \rangle)^3\rangle$ of the eigenvalues also follows a single-parameter scaling law
\begin{equation}
\label{eq:skew_g}
\frac{{\rm Skew}(\tilde{\tau})}{{\rm Skew}(\tilde{\tau}^{({\rm MP})})} = 1 + \frac{2 M_2}{g} + \frac{8M_2^2}{15g^2}
\end{equation}
once normalized by the eigenvalue skewness of an uncorrelated matrix, ${\rm Skew}(\tilde{\tau}^{({\rm MP})}) = \langle \tilde{\tau} \rangle^3 (M_1/M_2)^2$.
Again $M_2/g$ sets the departure from the uncorrelated behavior.
This scaling is validated with the experimental data in Fig.~2b, again with no free parameter.
In SI section III, we also calculate the MP-normalized fourth central moment (kurtosis) and find that in general it depends on two parameters $M_2/g$ and $M_2/M_1$, with the dependence on $M_2/M_1$ dropping out when it is small; 
this is validated with experimental data in Supplementary Fig.~3.

We also calculate the full eigenvalues distributions with the analytic FRM formalism.
For the parameters of the unknown matrix $t$ in our ansatz,
we use the measured average transmission $\bar{T} = 3\%$,
take $N_1 = g/\bar{T}$,
and $N_2 \approx 6 \times 10^5$
 as the number of CCD pixels when output collection is complete;
note that $N_1$ is not simply the number of modes in the illumination area because $g$ is enlarged by the lateral diffusion in an open geometry (SI section V).
The predicted eigenvalue distribution of the filtered matrix $\tilde{t}^\dagger \tilde{t}$ and its extremal values are obtained by solving equations (S8) and (S13) in SI section III.
These predictions, plotted in Figs.~3--4 as green solid lines, agree with the experimental data (green circles)
and differ from the uncorrelated MP law (gray dot-dashed lines).

We provide a simple heuristic model that approximates the FRM results.
The correlations reduce the degrees of freedom at the output, but have no effect on the input channels which 
are modulated independently by the $M_1$ SLM macropixels.
This suggests modeling the correlated $M_2$-by-$M_1$ matrix $\tilde{t}$ using an uncorrelated $M_2^{({\rm eff})}$-by-$M_1$ matrix $\tilde{t}^{({\rm eff})}$ with fewer output channels $M_2^{({\rm eff})} \le M_2$.
By choosing
\begin{equation}
M_2^{({\rm eff})} = M_2 \left( 1 + \frac{2 M_2}{3g} \right)^{-1},
\end{equation}
we match the eigenvalue variance of $\tilde{t}^{({\rm eff})}$ with that of $\tilde{t}$ given in equation~\eqref{eq:var_g}.
As expected, this effective degrees of freedom follows $M_2^{({\rm eff})} \approx M_2$ when $M_2 \ll g$, and $M_2^{({\rm eff})} \approx 3g/2$ when $M_2 \gg g$.
Within this model, 
all statistical quantities of $\tilde{t}^{({\rm eff})}$ follow the simple MP law with the reduced $M_2^{({\rm eff})}$, and all correlation effects are encapsulated in this reduction (which only depends on $M_2/g$).
For example, the correlation enhancement of control range
$\big( \tilde{\tau}_{\rm max}^{({\rm eff})} - \tilde{\tau}_{\rm min}^{({\rm eff})} \big) \big/ \big( \tilde{\tau}_{\rm max}^{({\rm MP})} - \tilde{\tau}_{\rm min}^{({\rm MP})} \big)$ plotted in Supplementary Fig.~2 is simply $\sqrt{ 1 +{2 M_2}/{3g} }$
when $M_2^{({\rm eff})} > M_1$.
For large targets ($M_2 \gg g$), the ratio $\tilde{\tau}_{\rm max/min}^{({\rm eff})}/ \langle \tilde{\tau}^{({\rm eff})} \rangle$ approaches $(1\pm\sqrt{2M_1/3g})^2$ and becomes independent of $M_2$,
meaning that the achievable total transmission enhancement (as studied in ref.~\onlinecite{2014_Popoff_PRL}) is determined solely by $M_1/g$ within this model.
Predictions of this ``effective MP model'' are reasonably good and are shown in Figs.~3--4 for the extremal eigenvalues and the eigenvalue distributions, in Supplementary Figs.~2--3 for the eigenvalue range and kurtosis.

The correlation-enhanced control enables more energy delivery into a multi-speckle-sized region, such as a photodetector for optical communications.
Meanwhile, the existence of non-local correlations~\cite{1987_Stephen_PRL, 1988_Feng_PRL, 1988_Mello_PRL} prevents the generation of truly independent random numbers through coherent diffusion~\cite{2014_Liutkus_srep}.
% 2016_Saade_ICASSP
These correlation effects can be enhanced or suppressed by changing the illumination spot size $D_{\rm in}$ (through the SLM macropixel size or through focus alignment~\cite{1997_Scheffold_PRB, 2013_Strudley_nphoton}), which changes $g$.
% 2014_Strudley_OL
Note that optimizing the target intensity does not necessarily optimize the target-to-background contrast, so imaging and photothermal therapy applications will require objective functions that account for both.
The theory here is applicable to microwave and acoustic wave, for which smaller $g$ can be readily achieved.
The correlations are even stronger when approaching localization~\cite{2014_Hildebrand_PRL},
and deep inside the localized regime ($g \ll 1$) the maximal transmission enhancement equals $M_1$ independent of the target size since a single eigenchannel dominates the transmission~\cite{2012_Shi_PRL, 2014_Pena_ncomms}.
Future work may extend the present formalism to broadband~\cite{2015_Hsu_PRL} and spatial-temporal control~\cite{2011_Aulbach_PRL, 2016_Mounaix_PRL}, and to the control of reflection and absorption in open disordered systems~\cite{2016_Liew_ACSPh, 2016_Yamilov_PRB, 2016_Yu_PRB}.
% 2011_Chong_PRL
% 2015_Choi_srep

\vspace{6pt}
\noindent {\bf Methods} \\
\noindent {\small Methods, including statements of data availability and any associated accession codes and references, are available in the online version of this paper.}

%\bibliography{references}
%\bibliographystyle{naturemag}

\vspace{8pt}
\noindent {\bf Acknowledgements} \\
\noindent {\small We thank S.~Popoff, Y.~Bromberg, S.~Knitter, R.~Sarma, W.~Xiong, F.~Scheffold, E.~Akkermans, and I.~M.~Vellekoop for helpful discussions, and the anonymous reviewers for their constructive comments.
This work is supported by the National Science Foundation under grant No.~DMR-1307632, DMR-1205307, and  ECCS-1068642, and by the US Office of Naval Research under grant No.~N00014-13-1-0649.
A.G.~acknowledges the support of LABEX WIFI (Laboratory of Excellence ANR-10-LABX-24) within the French Program ``Investments for the Future'' under reference ANR-10-IDEX-0001-02 PSL.}

\vspace{8pt}
\noindent {\bf Author Contributions} \\
\noindent {\small
C.W.H. and S.F.L. performed the experiment.
C.W.H. analysed the data.
C.W.H. developed the theory descriptions.
A.G. proposed the effective MP model.
H.C. and A.D.S. supervised the project.
All authors discussed and interpreted the results.
C.W.H. and A.D.S. wrote the manuscript with input from all authors.}
 
\vspace{8pt}
\noindent {\bf Additional information} \\
\noindent {\small
Supplementary information is available in the online version of the paper.
Reprints and permissions information is available online at www.nature.com/reprints.
Correspondence and requests for materials should be addressed to C.W.H. (chiawei.hsu@yale.edu).}

\vspace{8pt}
\noindent {\bf Competing financial interests} \\
\noindent {\small The authors declare no competing financial interests.}

\clearpage

\noindent {\bf Methods} \vspace{2pt} \\
\noindent {\small {\bf Experimental set-up.}
As illustrated in Fig.~1a, the expanded beam from a continuous-wave Nd:YAG laser (Coherent, Compass 215M-50 SL, wavelength $\lambda$ = 532 nm) is split into two parallel beams with orthogonal linear polarizations and equal intensity.
The two beams are modulated with different areas of a SLM (Hamamatsu, X10468-01) and then recombined.
Using a 4-f system (lenses L$_1$ and L$_2$; focal lengths = 21 cm), the surface of the SLM is imaged onto the entrance pupil of a microscope objective (NA$_{\rm in}$ = 0.95, Nikon CF Plan 100$\times$)
and then focused onto the scattering sample. 
The transmitted light is collected with an oil-immersion objective (NA$_{\rm out}$ = 1.25, Edmund DIN Achromatic 100$\times$), and the exit pupil of the objective is imaged onto a CCD camera (Allied Vision, Prosilica GC 660) through another 4-f system (lenses L$_3$ and L$_4$; focal lengths = 20 cm).
In this way, the SLM pixels and the CCD pixels correspond to the different angles incident onto and transmitted from the sample.

The nearby SLM pixels are grouped into square macropixels; smaller macropixels correspond to more finely spaced incident angles, which yields a larger illumination spot and provides more input channels (macropixels) for modulation.
For example, when one macropixel consists of 8$\times$8 SLM pixels, we have $M_1^{\rm (tot)} = 978$ macropixels (489 per polarization) imaged onto the entrance pupil of the input objective and available to use.
We measure the transmission matrix for the $M_1 = 512$ input channels (macropixels) at the center (16$\times$16 per polarization), using the other $M_1^{({\rm ref})} = 466$ available macropixels as the reference.
Then we synthesize the desired wavefront at the $M_1$ macropixels; at this stage the other macropixels are ``switched off'' by displaying a high-spatial-frequency phase pattern, making them blocked by the iris placed at the Fourier plane of the lens L$_1$.

For the theoretical prediction of the dimensionless conductance $g$ (SI section V), we need to know the size and spatial profile of the illumination spot.
In our set-up, the SLM macropixels are mapped to the space $(k_x, k_y)$ of transverse wavevectors on the surface of the sample. Let $q \times q$ denote the size of one macropixel in the $k$ space. Then the intensity profile of the incident light is
\begin{equation}
\label{eq:I_sinc_3D}
I(x,y) \sim {\rm sinc}^2({qx}/{2}) {\rm sinc}^2({qy}/{2}).
\end{equation}
The illumination area is $A_{\rm in} \equiv [\iint dxdy I(x,y) ]^2/\iint dxdy I^2(x,y) = (3\pi/q)^2$.
The available number of macropixels is the number of $q \times q$ squares within two circles of radii $(2\pi/\lambda) {\rm NA}_{\rm in}$ in the $k$ space, so $M_1^{\rm (tot)} = 2\pi (2\pi/\lambda)^2 ({\rm NA}_{\rm in})^2/q^2$.
Thus we can determine the area $A_{\rm in}$ and the diameter $D_{\rm in} \equiv \sqrt{4A_{\rm in}/\pi}$ of the illumination spot from $M_1^{\rm (tot)}$ and $\lambda$.
In this study we use 16$\times$16-, 8$\times$8-, and 4$\times$4-sized macropixels, corresponding to
$M_1 = 128, 512, 2048$
and $D_{\rm in} \approx 6, 12, 24$ $\mu$m.

\vspace{6pt}
\noindent {\bf Sample preparation and characterization.}
We dissolve 6 grams of ZnO microparticle powder (Alfa Aesar Puratronic) into 3 mL of deionized water and 3 mL of ethanol, and sonicate the mixture in an ultrasonic bath for 30 minutes for thorough dispersion.
The solution is then spin-coated onto a microscope cover slip (22$\times$22 mm$^2$, thickness 0.15 mm) and allowed to dry.
The resulting sample is shown in the inset of Fig.~1a.
Scanning electron microscopy shows that the ZnO particle diameters center around 200 nm.
We use the center of the sample, where the thickness of the ZnO film is measured by a profilometer to be $L \approx$ 60 $\mu$m, and the total transmission is measured with a photodetector to be $\bar{T} \approx$ 3\%.

\vspace{6pt}
\noindent {\bf Transmission matrix measurement.}
We measure the transmission matrix $\tilde{t}$ using a modified version of the phase-shifting common-path interferometric method.
As in refs.~\onlinecite{2010_Popoff_PRL, 2011_Popoff_NJP}, we control $M_1$ macropixels on the SLM and switch on additional $M_1^{({\rm ref})}$ macropixels with a flat phase as the reference.
When the $a$-th input mode is sent in with a relative phase of $\phi$, the measured intensity on the $b$-th pixel of the CCD is $|e^{i\phi} \tilde{t}_{ba} + s_b|^2$, where $s_b$ denotes the transmitted field from the reference pixels.
With four measurements at $\phi = 0, \pi/2, \pi, 3\pi/2$, we obtain $u_{ba} = \tilde{t}_{ba} s_b^*$.
We perform the measurements with the input channels $\{a\}$ in the Hadamard basis, and then transform back to the basis of SLM macropixels ({\it i.e.} incident angles).

So far this method only yields $u_{ba}$, which is the transmission matrix $\tilde{t}_{ba}$ multiplied by an unwanted field $s_b^*$ from the reference, which obscures the relative phase and amplitude between the output channels $\{b\}$.
To study $\tilde{t}^\dagger \tilde{t}$ and the transmitted intensity, it is necessary to recover the relative amplitude between $\{b\}$.
Thus, for each input channel $a$, we perform an additional measurement with the $M_1^{({\rm ref})}$ reference pixels switched off, which provides $|\tilde{t}_{ba}|^2$ that contains the relative amplitude between the output channels.
Specifically, we use $u_{ba}|\tilde{t}_{ba}|/|u_{ba}| = \tilde{t}_{ba} s_b^*/|s_b|$ as our transmission matrix; the relative phase between $\{b\}$ is still unknown but is irrelevant for us.
The whole measurement process takes 2, 8, 32 minutes respectively for $D_{\rm in} \approx 6, 12, 24$ $\mu$m.

For each illumination diameter, we measure 10 transmission matrices at sufficiently different times that there is no discernible correlation between the measured matrices. The 10 sets of data are used to improve the statistics of the eigenvalues and the intensity correlations. 

\vspace{6pt}
\noindent {\bf Determination of optimal phase-only wavefront.}
With the transmission matrix measured, we can instantly determine the optimal wavefront for any given target.
When both amplitude and phase can be modulated, the optimal wavefront for maximal [minimal] transmission into the target is simply the eigenvector of $\tilde{t}^\dagger \tilde{t}$ with the largest [smallest] eigenvalue;
this is not the case for phase-only modulation.
For maximal transmission of a phase-only wavefront, we use the phase part of the maximal eigenvector.
For minimal transmission, taking the phase of the minimal eigenvector is not optimal (for reasons explained in ref.~\onlinecite{2016_Sarma_PRL}), so we look for the phase profile $\{\phi_1, \ldots, \phi_{M_1}\}$ that minimizes $\sum_{b=1}^{M_2} \left| \tilde{t}_{ba} e^{i\phi_a} \right|^2$,
using the gradient-based low-storage BFGS algorithm$^{51,52}$ implemented in the free optimization package NLopt$^{53}$.
The optimization only takes a few seconds.

\vspace{6pt}
\noindent {\bf Data availability.}
The data that support the plots within this paper and other findings of this study are available from the corresponding author upon reasonable request.
}

\vspace{8pt}
\noindent {\bf References} \\
\vspace{-7pt}
{\small
\begin{enumerate}
\setcounter{enumi}{50}
\item Nocedal, J. Updating quasi-Newton matrices with limited storage. {\it Math. Comput.} {\bf 35}, 773?782 (1980).
\item Liu, D. C. \& Nocedal, J. On the limited memory BFGS method for large scale optimization. {\it Math. Program.} {\bf 45}, 503?528 (1989).
\item Johnson, S. G. The NLopt nonlinear-optimization package, http://ab-initio.mit.edu/nlopt.
\end{enumerate}
}

\end{document}